\documentclass[%
 reprint,
amsmath,amssymb,
pra,
]{revtex4-2}

\usepackage{graphicx}
\usepackage{dcolumn}
\usepackage{bm}

\begin{document}

\preprint{APS/123-QED}

\title{High-Order Mie Resonance and Transient Field Enhancement in Laser-Driven Plasma Nanoshells}

\author{Xiaohui Gao}
  \email{gaoxh@utexas.edu}
\affiliation{Department of Physics, Shaoxing University, Shaoxing, Zhejiang 312000, China}

\date{\today}

\begin{abstract}
We demonstrate substantial field enhancement in plasma nanoshells through high-order Mie resonances using combined Mie theory and particle-in-cell simulations. Optimal shell geometries yield approximately threefold electric field enhancement for 800 nm irradiation, with transient buildup times of tens of femtoseconds before plasma expansion disrupts resonance. Few-cycle pulses produce reduced enhancement due to insufficient resonance establishment. These findings enable optimized laser-plasma interactions for applications including diagnostics of laser-cluster interaction and energetic ion production from engineered core-shell targets, highlighting the critical role of temporal dynamics in nanoplasma resonances.
\end{abstract}
\maketitle
\section{Introduction}
Resonance is perhaps the most compelling phenomenon in physics, underpinning a vast range of applications from laser technology to medical imaging. In high-field physics, resonant mechanisms are responsible for a host of fascinating processes, including resonant absorption in laser-plasma interactions~\cite{Kruer1988,Leshchenko2023} and the dramatic enhancement of nonlinear optical effects through plasmon excitation~\cite{Vampa2017NP}.

In the context of laser interactions with finite particles, resonance occurs when the laser frequency matches a natural eigenfrequency of the system. For a bulk plasma, this characteristic frequency is the plasma frequency $\omega_p=\sqrt{n_e e^2/\epsilon_0 m_e}$, where $e$ is the elementary charge, $\epsilon_0$ is the vacuum permittivity, and $m_e$ is the electron mass. At a typical laser wavelength of $800$~nm, this corresponds to a critical electron density of $n_c \approx 1.7\times10^{21}~\text{cm}^{-3}$, which is orders of magnitude lower than solid density. The natural frequency exhibits significant geometry dependence: for a spherical particle in the Rayleigh regime, the fundamental dipolar surface plasmon resonance redshifts to $\omega_p/\sqrt{3}$. This geometry-induced resonance plays a crucial role in laser-cluster interactions~\cite{Fennel2007PRLa}. Consequently, efficient laser-plasma coupling in solid-density targets typically requires hydrodynamic expansion to reach the resonant density profile~\cite{Ditmire1997PRL}, specialized dual-pulse schemes to create the resonant structure~\cite{Doppner2005PRL}, or nonlinear absorption pathways~\cite{Kundu2006PRL, Mulser2005PRA}.

Structured targets offer an alternative paradigm for engineering resonant conditions without requiring large-scale plasma expansion. Core-shell geometries, widely investigated in nanophotonics, support unique plasmonic modes arising from the hybridization of sphere and cavity plasmons~\cite{Prodan2003S}. In such systems, field enhancement occurs in some regions at the expense of field suppression in some other regions~\cite{Neeves1989JOSAB}, with the enhancement factor exhibiting strong dependence on the electron collision frequency~\cite{Averitt1999JOSAB}. 

Here we investigate intense laser interactions with subwavelength plasma shells. In our study, we consider a fully ionized hydrogen nanoshell, modeled as the ionized skin layer of a cluster. Hydrogen clusters are commonly employed in proton acceleration studies~\cite{Jinno2018PPCF, Aurand2019PP} and require only moderate laser intensities to achieve full ionization. The shell geometry is defined by inner radius $r_a$, outer radius $r_b$, and thickness $\delta=r_b-r_a$. A physically analogous scenario of transient plasma nanoshell formation occurs naturally during the intense laser irradiation of large clusters, where the periphery is rapidly ionized while the core remains neutral, effectively creating an ionized nanoshell~\cite{Liseykina2013PRL, Fukuda2007PLA}. Probing such systems with weaker secondary pulses could yield valuable insights into laser-cluster interaction dynamics. Alternatively, fabricated hollow carbon nanospheres~\cite{Li2016JMCA} and micron-sized gas bubbles~\cite{Dollet2019} present experimentally accessible platforms for controlled studies. These structures can generate enhanced local fields exceeding the ionization threshold across significant volumes. Our study differs from low-intensity metallic nanoshell systems due to the involvement of ionization dynamics, rapid heating, and plasma expansion. We particularly focus on large shell structures where high-order Mie resonances can be excited, resulting in substantial volumes of enhanced field.

This paper is organized as follows. In Sec.~\ref{sec:mie}, we present a theoretical investigation of the linear optical response of nanoshells using Mie theory, identifying the conditions for exciting high-order Mie resonances. In Sec.~\ref{sec:pic}, we explore the non-equilibrium dynamics of intense laser-nanoshell interactions using particle-in-cell (PIC) simulations. This time-dependent analysis is crucial, as the transient effects of pulsed laser excitation and rapid plasma expansion significantly alter the interaction dynamics compared to the continuous-wave assumption of Mie theory. Finally, Sec.~\ref{sec:conclusion} provides a summary of our principal findings and discusses their implications.

\section{Mie Calculation}
\label{sec:mie}
The electromagnetic fields inside and around the shell under the irradiation of a continuous plane wave linearly polarized in the $y$ direction are computed using the Mie scattering code Scattnlay v2.0~\cite{Ladutenko2017CPC}. The code numerically evaluates the analytical solution expressed as an infinite series of scattering coefficients, which are derived from Maxwell's equations subject to the boundary conditions at the material interfaces.

We assume a hollow core configuration, noting that while a dielectric core would induce a slight resonance shift, it would not qualitatively alter our results. The dielectric response of the plasma shell is described by the Drude model:
\begin{equation}
\epsilon_c(\omega) = 1 - \frac{\omega_p^2}{\omega^2(1 + i\nu_c/\omega)},
\label{eq:drude}
\end{equation}
where $\omega$ is the laser angular frequency, and $\nu_c$ represents the collision frequency. The atomic density is taken as that of liquid hydrogen, $n_{\text{atom}} = 4.56 \times 10^{22}\,\text{cm}^{-3}$~\cite{Aurand2019PP}, yielding a plasma frequency of $\omega_p = 1.2 \times 10^{16}$ rad/s upon complete ionization. The collision frequency is typically on the order of 1 fs$^{-1}$ for dense plasmas~\cite{Moll2010JPBAMOP} including surface collision contributions~\cite{Megi2003JPB}. We also examine the sensitivity of our results to variations in $\nu_c$.

The plasma shell thickness is fixed at 20 nm throughout our analysis. For $\lambda = 800$ nm and $\nu_c = 1$ fs$^{-1}$, the skin depth in the hydrogen plasma is approximately 25 nm calculated using the formula for a collisional plasma~\cite{Chen2001PP}, ensuring that our chosen thickness is physically relevant. In the subsequent calculations, we maintain this fixed shell thickness while systematically varying the inner radius $r_a$ to explore the parameter dependence of the resonant response.

\begin{figure}[htbp]
\centering
\includegraphics[width=0.48\textwidth]{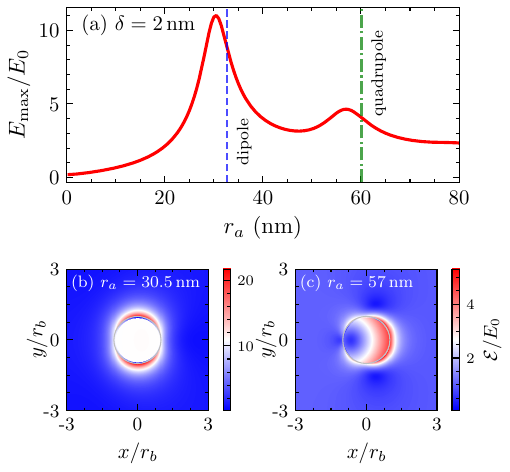}
\caption{Electric field enhancement in core-shell nanoparticles under quasi-static conditions. (a) Maximum field enhancement factor ($E_{\text{max}}/E_0$) as a function of core radius $r_a$ for a nanoshell with 2 nm thickness. Vertical lines indicate the dipole (32.7 nm, blue) and quadrupole (60.1 nm, green) resonance positions predicted by analytical theory. (b,c) Electric field distributions ($E/E_0$) in the $xy$ plane at resonance conditions: (b) dipole resonance at $r_a = 30.5$ nm and (c) quadrupole resonance at $r_a = 57$ nm.}
\label{shell-plot2}
\end{figure}

As a benchmark for the Mie calculations, Fig.~\ref{shell-plot2} presents results for a nanoshell of thickness $\delta = 2$ nm with collision frequency $\nu_c = 0.2$ fs$^{-1}$. The red curve shows the maximum field enhancement within the nanoparticle volume ($r < r_b$), revealing distinct resonances at 30.5 nm and 57 nm. The resonant frequencies for a spherical plasma shell are given by the plasmon hybridization model in the electrostatic approximation~\cite{Prodan2003S}:
\begin{equation}
   \omega_{l\pm}^2 = \frac{\omega_p^2}{2} \left[ 1 \pm \frac{1}{2l+1} \sqrt{1 + 4l(l+1)\left(\frac{r_a}{r_b}\right)^{2l+1}} \right], 
\label{eq1}
\end{equation}
where $l$ denotes the multipole order. We consider the lower frequency branch so that resonance is possible at solid density for 800-nm. The analytical model predicts resonance inner radii of approximately 32.7 nm for the dipole mode ($l=1$) and 60.1 nm for the quadrupole mode ($l=2$), indicated by vertical lines in Fig.~\ref{shell-plot2}(a). The Mie calculation shows a slight shift in the resonance positions since Eq.~\ref{eq1} neglects collisional damping. The corresponding field distributions in Figs.~\ref{shell-plot2}(b) and ~\ref{shell-plot2}(c) clearly demonstrate the characteristic dipole and quadrupole patterns, respectively.

\begin{figure}[htbp]
\centering
\includegraphics[width=0.40\textwidth]{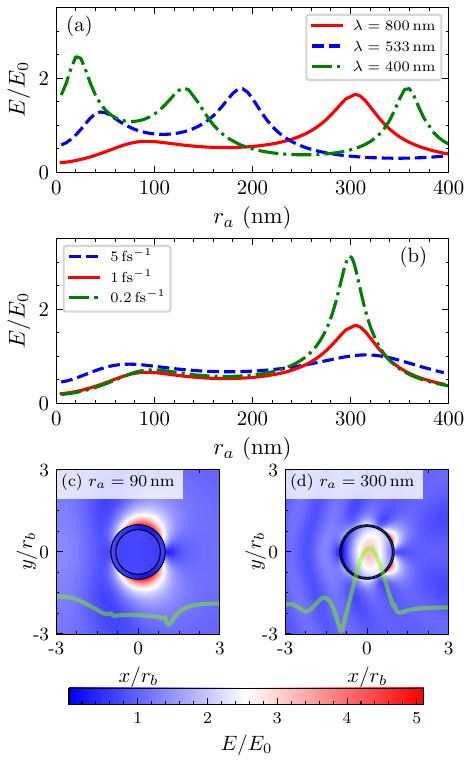}
\caption{Electric field enhancement in hydrogen nanoshells with 20 nm thickness. (a) Field strength at cluster center versus inner radius $r_a$ for three driving wavelengths: 800 nm (red), 533 nm (blue), and 400 nm (green), with $\nu_c = 1$ fs$^{-1}$. (b) Field strength dependence on collision frequency: $\nu_c = 0.2$ fs$^{-1}$ (green), 1 fs$^{-1}$ (red), and 5 fs$^{-1}$ (blue). (c) Electric field distribution in the x-y plane for $r_a = 90$ nm, with a lineout at $y=0$ (green curve). (d) Field distribution for $r_a = 300$ nm.}
\label{shell-plot}
\end{figure}

Figure~\ref{shell-plot}(a) displays the electric field at the cluster center as a function of inner radius for a 20 nm plasma shell with vacuum core. The red, blue, and green curves correspond to driving wavelengths of 800 nm, 533 nm, and 400 nm, respectively, with $\nu_c = 1$ fs$^{-1}$. As the laser frequency increases, the first resonance shifts to smaller core sizes while its amplitude increases. The first resonance at 400 nm satisfies the quasi-static approximation, showing good agreement with analytical predictions. In contrast, the 800 nm case lies outside this regime. Multiple resonances are observed across all wavelengths. The resonant enhancement exhibits strong sensitivity to collision frequency, as shown in Fig.~\ref{shell-plot}(b). For $\lambda=800\,$nm and $\nu_c = 0.2$ fs$^{-1}$, we observe a pronounced field enhancement factor of three at $r_a=300$ nm, consistent with PIC simulation results presented in the following section. A minor peak appears at 90 nm. The corresponding field distributions in Figs.~\ref{shell-plot}(c) and \ref{shell-plot}(d) exhibit dipole and multipole characteristics, respectively. The resonant behavior in the Mie scattering regime has been analyzed for thin and thick shell limits~\cite{Tzarouchis2018ITAP}, where inhomogeneous polarization and retardation effects enable excitation of higher-order modes~\cite{Ahmed2019MRE}.

\section{PIC Simulations of Transient Dynamics}
\label{sec:pic}

While Mie scattering calculations provide valuable insight into field distributions within uniform nanoshells under continuous-wave excitation, they involve several simplifying assumptions that limit their applicability to intense, pulsed laser interactions. The continuous-wave framework assumes steady-state conditions, neglecting transient dynamics inherent to finite-duration laser pulses. Additionally, the use of a constant, spatially uniform dielectric function may not fully capture the non-equilibrium plasma response under intense irradiation.

To overcome these constraints and capture the realistic interaction physics, we conduct three-dimensional particle-in-cell computations employing the open-source framework \textsc{Smilei}~\cite{Derouillat2018CPC}. Our computational domain extends over $8\lambda \times 3\lambda \times 3\lambda$, with $\lambda = 800$ nm. The spatial resolution is configured at $\lambda/80$ across all dimensions, with domain dimensions selected to guarantee solution convergence. Temporal discretization uses $T/145$ increments, where $T \approx 2.67$ fs represents the laser period, maintaining a Courant number of 0.956 to ensure numerical stability. A hydrogen nanoshell is centered within the simulation volume, with a 20 nm wall thickness unless otherwise specified. Initial conditions incorporate 64 neutral macro-particles per cell without pre-existing free electrons.

To investigate transient resonance dynamics, we employ a trapezoidal laser pulse profile with a three-cycle linear ramp followed by a plateau, with peak intensity of $1.5\times10^{15}$ W/cm$^2$. This intensity ensures rapid field ionization while maintaining relatively slow plasma expansion. The uniform transverse electromagnetic wave propagates along the positive x-axis with linear polarization aligned to the $y$-direction, incident from the left domain boundary. Computational boundaries apply periodicity in transverse coordinates while implementing Silver-Muller conditions along the longitudinal direction ($x$-axis) for electromagnetic components. Particles intersecting longitudinal boundaries undergo removal from the simulation. Electron-ion and electron-electron collisions are enabled with the module using a Monte Carlo algorithm, though collisional ionization is omitted since field ionization dominates plasma formation under these high-intensity conditions, leading to rapid full ionization on femtosecond timescales.

\begin{figure}[htbp]
\centering
\includegraphics[width=0.45\textwidth]{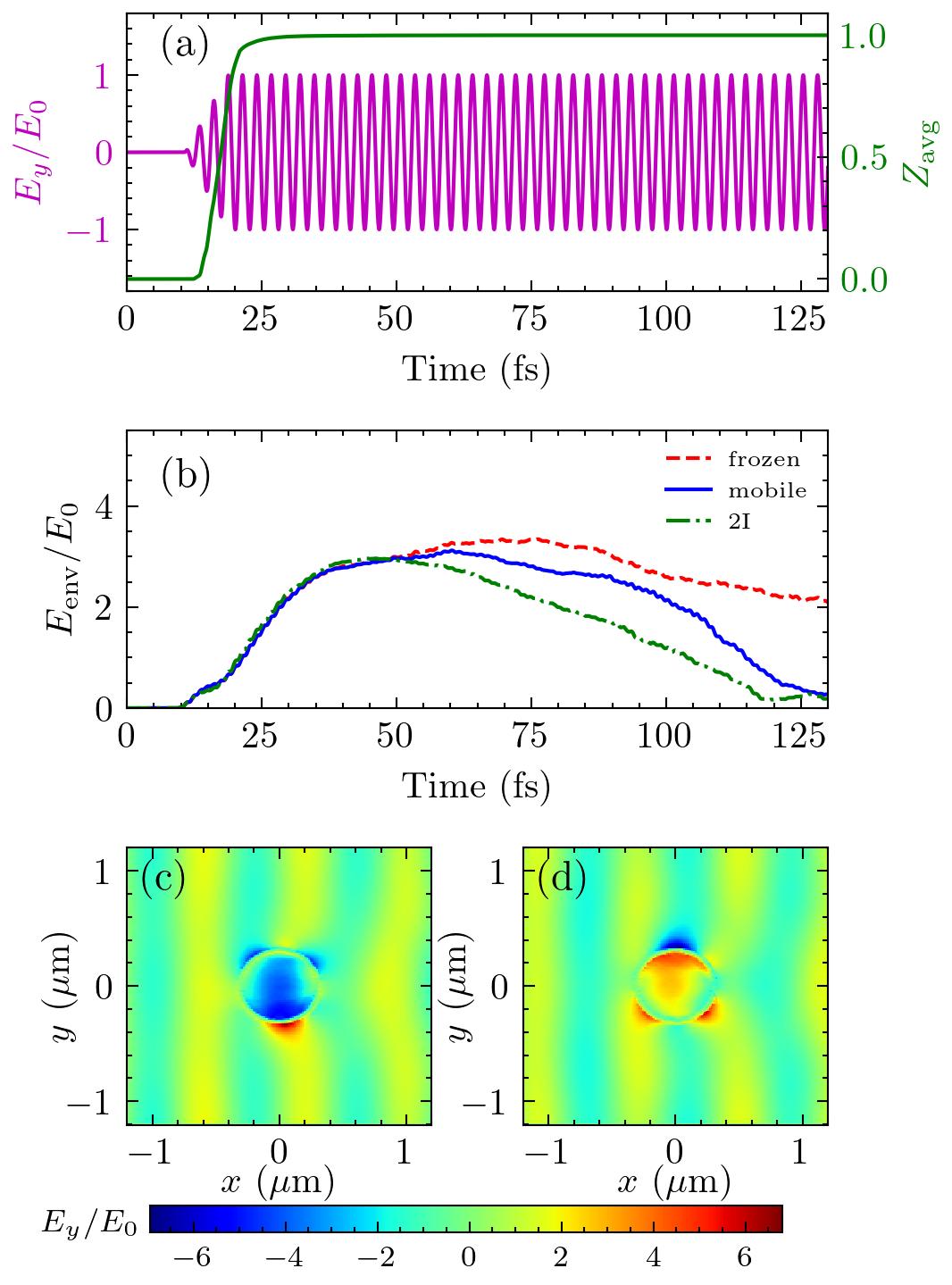}
\caption{Transient dynamics of laser-nanoshell interaction for $r_a = 300$ nm hydrogen nanoshell. (a) Normalized electric field in vacuum (magenta line) on left axis, with average charge state (green line) on right axis. (b) Field envelope evolution at nanoshell center for three cases: frozen ions (red dashed), mobile ions (blue solid), and mobile ions with doubled laser intensity (green dash-dotted). (c,d) Electric field profiles in the polarization plane at (c) 48.80 fs and (d) 52.48 fs.}
\label{pic1}
\end{figure}

Figure~\ref{pic1} illustrates ionization dynamics and field enhancement for a 300 nm inner radius nanoshell. The vacuum laser field (magenta line in Fig.~\ref{pic1}a) and field at the nanoshell center (blue line) are shown alongside the average charge state (green dashed line), which indicates nearly complete ionization following the pulse ramp. Figure~\ref{pic1}(b) displays the field envelope evolution for three scenarios: frozen ions (red dashed), mobile ions (blue solid), and mobile ions with doubled intensity (green dash-dotted). As the blue curve shows, the field reaches a maximum enhancement factor of approximately three with a buildup time of $\sim$ 25 fs, sustained for about 80 fs. In a simple harmonic oscillator model, the buildup time $\tau\approx2/\nu_c$. Based on Mie calculations, the equivalent $\nu_c \approx 0.2$ fs$^{-1}$ yields a buildup time of 10 fs. The actual buildup time is also influenced by the three cycles slope of the driving pulse, so the expected value agrees with the observed one. The independence of buildup time from laser field is confirmed by the doubled intensity case. However, the enhancement decay is faster because hotter electrons expands faster. Field enhancement diminishes due to plasma expansion, which modifies the resonant condition by altering the density profile.  With frozen ions, the minor enhancement decrease after 100 fs may result from collision frequency variations and hot electron streaming. Figures~\ref{pic1}(c) and ~\ref{pic1}(d) show polarization plane field snapshots at 48.80 fs and 52.48 fs, respectively, revealing multipole field patterns with significant enhancement over large volumes. Note that Fig.~\ref{pic1}(d) and Fig.~\ref{shell-plot}(d) cannot be compared directly as the former is the field at a given time and the latter is the amplitude distribution in a steady state. This transient behavior underscores the importance of temporal dynamics in laser-nanoshell interactions, where optimal field enhancement occurs during a brief window between resonance buildup and plasma expansion-induced disruption.

\begin{figure}[htbp]
\centering
\includegraphics[width=0.40\textwidth]{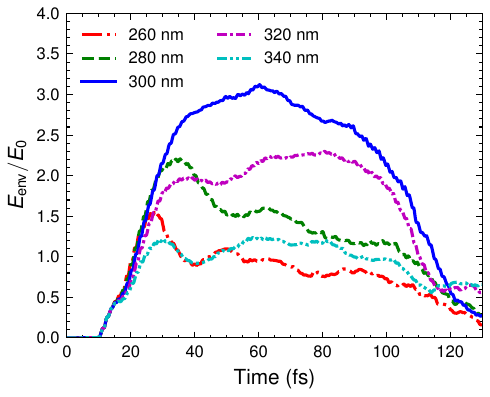}
\caption{Temporal evolution of electric field envelopes for hydrogen nanoshells with different inner radii (260-340 nm). }
\label{pic2}
\end{figure}

Field enhancement across different nanoshell sizes is shown in Fig.~\ref{pic2}. Optimal enhancement occurs at 300 nm inner radius, consistent with Mie theory predictions in Section~\ref{sec:mie}. This agreement validates our simulation approach. For radii between 260 nm and 340 nm, enhancement persists for approximately 100 fs, which is likely due to similarity in the expansion dynamics. Smaller shells exhibit earlier maximum enhancement, while larger shells exhibit delayed maximum enhancement, due to the evolution of the density profile.

\begin{figure}[htbp]
\centering
\includegraphics[width=0.35\textwidth]{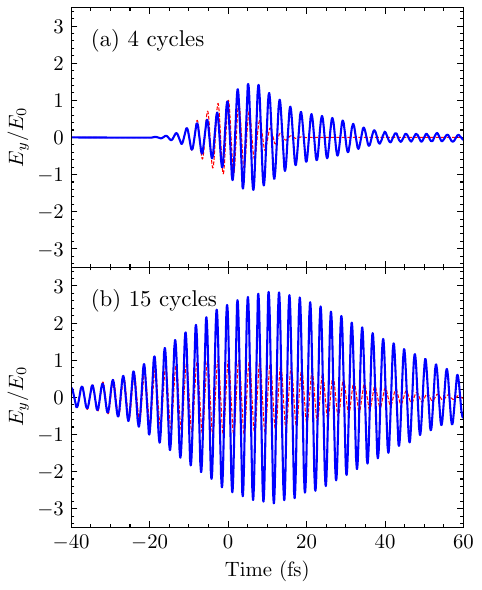}
\caption{Time evolution of transverse electric field $E_y$ in laser-plasma interactions. (a) Field component $E_y$ for 4-cycle FWHM pulse (Blue solid line). Red dotted line shows field without nanoshell. (b) Similar comparison for 15-cycle FWHM pulse.}
\label{pic3}
\end{figure}

Figure~\ref{pic3} shows field evolution under 4-cycle and 15-cycle FWHM (intensity) pulses. For these simulations, the simulation domain is increased to $8\lambda \times 4.5\lambda \times 4.5\lambda$ to ensure convergence. A larger transverse domain is needed likely due to the drift velocity under the irradiation of a few-cycle pulse. The spatial grid size is set to $\lambda$/64 to to accommodate computational resource constraints. Both cases exhibit delayed field maxima due to resonance buildup time, with the first delayed by 5 fs and the second subplot delayed by 10 fs. For the 4-cycle pulse, the enhancement is only 1.5. Achieving full enhancement typically requires pulses of tens of cycles. The damping of electron motion in the plasma naturally results in a buildup of the resonance and produces a delayed maximum response. Static calculations such as Mie theory would overestimate the resonant field enhancement driven by few-cycle pulses.

\section{Conclusion}
\label{sec:conclusion}
We demonstrate substantial field enhancement in plasma nanoshells through excitation of high-order Mie resonances. Mie theory and PIC simulations identify consistent optimal resonance conditions at specific shell geometries, yielding approximately threefold electric field enhancement (corresponding to nearly an order of magnitude intensity enhancement) for 800 nm laser irradiation. The transient nature of this enhancement is characterized by resonance buildup times of tens of femtoseconds, followed by rapid disruption due to plasma expansion. Crucially, pulse duration significantly influences the amplification, with few-cycle pulses producing reduced enhancement due to insufficient time for resonance establishment—a universal limitation for few-cycle-driven resonances.

Our results underscore the critical role of temporal dynamics in nanoplasma resonances. These findings suggest promising applications in laser-matter interactions. For transient plasma shells with unionized cores, enhanced probe pulses could enable detailed investigation of laser-cluster dynamics through harmonic generation. Furthermore, similar enhancement mechanisms could be realized in hollow carbon nanospheres at 400 nm wavelength with appropriately tuned dimensions. When combined with nanometer-sized metallic particles in the hollow core, such field-enhanced systems may facilitate energetic ion generation via Coulomb explosion. 

\begin{acknowledgments}
This work was supported by Natural Science Foundation of Zhejiang Province (LY19A040005).
\end{acknowledgments}

\section*{Data Availability}
The data that support the findings of this study are available from the corresponding author upon reasonable request.

%

\end{document}